\documentclass[10pt,letterpaper,onecolumn]{article}
\usepackage{amsmath,amssymb,graphicx,cite,url,float,dsfont}

\begin{document}
\title{\textsf{\textbf{Domino successive-deviation ghost imaging}}}

\author{Ya-Xin Li$^{1,2,3}$, Wen-Kai Yu$^{1,2,3,4}$, Shuo-Fei Wang$^{1,2}$}

\footnotetext[1]{Center for Quantum Technology Research, School of Physics, Beijing Institute of Technology, Beijing 100081, China}
\footnotetext[2]{Key Laboratory of Advanced Optoelectronic Quantum Architecture and Measurements of Ministry of Education, School of Physics, Beijing Institute of Technology, Beijing 100081, China}
\footnotetext[3]{These authors contributed equally to this work}
\footnotetext[4]{Correspondence and requests for materials should be addressed to W.-K.Yu (email: yuwenkai@bit.edu.cn)}

\date{}

\maketitle

\renewenvironment{abstract}{%
    \setlength{\parindent}{0in}%
    \setlength{\parskip}{0in}%
    \bfseries%
    }{\par\vspace{12pt}}

\begin{abstract}
Traditional ghost imaging acquires images via the correlation of the intensity fluctuations of reference patterns and bucket values, and can even generate positive-negative images by conditionally averaging partial patterns. Here, we propose a domino successive-deviation ghost imaging method, which owns a good image quality comparable to that of differential ghost imaging with real-time fast computation, and a better robustness in practical scenarios where measurement noise and the instability of illuminating source coexists. Furthermore, it happens to generate real-time positive and negative images, giving a new insight into physical essence of positive-negative ghost image phenomenon. Both simulation and experimental results have demonstrated the feasibility of our approach. Therefore, this work complements the theory of ghost imaging and opens a door to practical applications of real-time single-pixel imaging.
\end{abstract}

\noindent The emergence of single-pixel imaging (SPI) \cite{Matthew2019,BSun2013,SZhang2017} provides possibilities for many scenarios where pixelated array detectors are unavailable, and reduces the need for cumbersome and expensive camera lenses. Hadamard \cite{MJSunNC2016} and Fourier \cite{Zhang2015} SPI schemes use complete orthogonal basis for sampling and matrix inversion (or multiplication) for reconstruction, but with sensitivity to noise \cite{ZibangZhang2017,YuOC2017}. As an alternative, ghost imaging (GI) \cite{Pittman1995,Bertolotti2019} can acquire the object image via second-order intensity correlation of reference patterns and single-pixel (bucket) values, which is much faster than matrix multiplication. To our knowledge, GI was primitively experimentally demonstrated by utilizing a biphoton source \cite{Pittman1995}, later it was also realized with thermal light \cite{Gatti2004,DaZhang2005} or pseudo-thermal light \cite{Jun2005}. Although GI has an innate advantage in computation, its imaging quality is not very high due to its statistical average nature.

In order to improve the image quality, many GI algorithms \cite{YuOE2014,Yao2014} have been proposed, such as the approach of removing the background item ($\Delta$GI) \cite{Gatti2004}, differential ghost imaging (DGI) \cite{Ferri2010}, and the like. Among all these, DGI is nearly the best correlation function. To acquire a high-visibility, a large number of measurements are required to suppress the background noise, which leads to long acquisition time, huge computation overhead and large memory consumption. For example, using X-ray GI \cite{Hong2016,Xin2018} to observe living organisms, the long-time sampling will cause irreversible damages to the specimens. Recently, a GI method named correspondence imaging (CI) was put forward \cite{Luo2012,Wen2012,YuCPB2015,MJSunAO2015} to improve the visibility by generating positive and negative images from conditioned partial patterns. Although it also fails to reduce the acquisition time (exposure number), it brings a dawn to further reduce the number of measurements. Moreover, it is hard for all above approaches to work properly under extreme measurement environments.

In this paper, based on previous work of complementary/differential measurements \cite{YuSR2014,WenKai2015,Yu2015,YuOC2016}, a method named successive-deviation GI (SGI) is presented. The main idea is to make a difference of two adjacent bucket values or random patterns (the $(i+1)$th bucket-value/pattern minus the $i$th one, vice versa, this successive-deviation process is just like dominoes), where double measurements of a complementary pattern pair (one pattern followed by its inversed one) are no longer necessary \cite{BSun2013,WenKai2015,Yu2015}. Unlike DGI \cite {Ferri2010}, our scheme takes full advantage of shift-multiplexing-based successive-deviation, without the need to subtract the ensemble average, the noise variance can be greatly averaged/suppressed (especially for double-arm GI with a rotating ground glass) and the images can be recovered in real-time. The technique can obtain a good image quality comparable to that of DGI, and even better under conditions where temperature drift of light source and noise coexists. The variants of our method can also generate positive and negative images, but does not need to sample all the patterns to get an average before reconstruction. We have verified the feasibility of this technique via both numerical simulation and classic double-arm lensless GI setup using a pseudo-thermal source.

\begin{figure}[htbp]
\centering
\includegraphics[width=\linewidth]{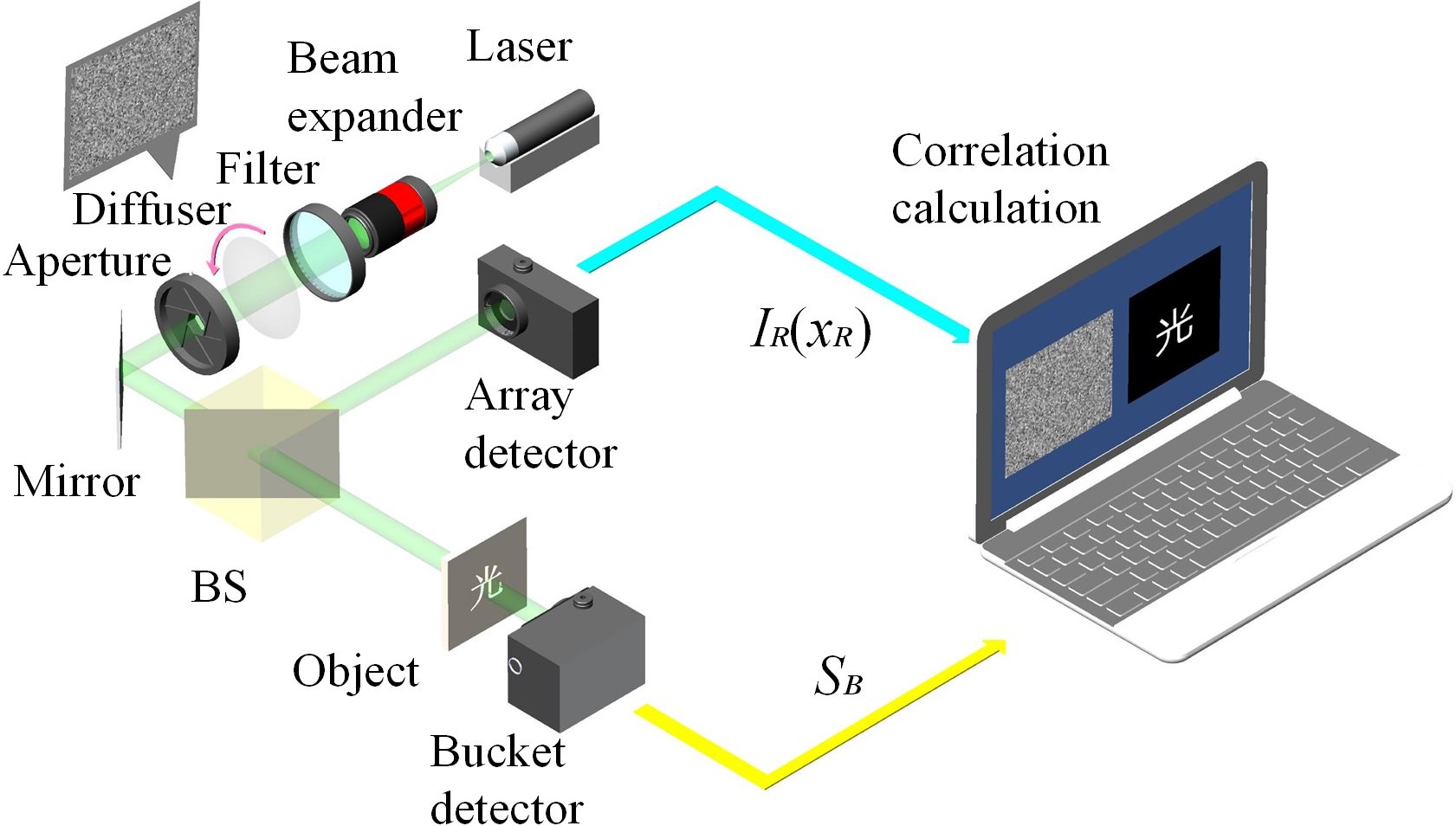}
\caption{Schematic diagram of double-arm lensless GI setup.}
\label{fig:Setup}
\end{figure}

Our experiment is based on a classic double-arm lensless GI setup, as shown in Fig.~\ref{fig:Setup}, where the semiconductor laser light of wavelength $\lambda=532$~nm, with its beam being expanded to 1.92~mm diameter, passes through a diffuser (i.e., a ground-glass disk rotating at 0.3~rad/s controlled by a stepper motor) to produce pseudothermal light. The role of the beam expander is to make the laser light illuminate a larger region of the ground glass to obtain relatively finer speckles. The light field of time-varying random speckles passes through an aperture and is reflected by a mirror to a 50:50 beam splitter (BS), which divides the light into two arms. One arm (reference arm) merely records the light field distribution $I_R(x_R)$ of the source by a charge-coupled-device (CCD), i.e., array detector, and another (object arm) collects the total intensity from the transmitted object with a single-pixel (bucket) detector, denoted by $S_B$. The space coordinate $x_R$ refers to the spatial positions in the reference arm, and can be stretched into one-dimensional signal for simplicity. After $N$ measurements, we can reconstruct the object images via intensity correlations with help of computer.

In GI, the object image can be retrieved via second-order correlation of the reference signal $I_R(x_R)$ and the bucket signal $S_B$:
\begin{equation}
G^{(2)}=\left\langle S_BI_R(x_R)\right\rangle,
\label{eq:refname1}
\end{equation}
where $\left\langle u\right\rangle=\frac{1}{N}\sum_{i=1}^Nu^i$ denotes an ensemble average of the signal $u$. $S_{B}=\int_{A_l}I_B(x_B)T(x_B)dx_B$, where $x_B$ and $I_B$ is the spatial coordinate and the light field of the object beam, respectively, $T(x_B)$ denotes the transmission function of the object, $A_l$ stands for the integration area. By deducting the ensemble average items from $S_B$ and $I_R(x_R)$, then we will have
\begin{align}\label{eq:refname2}
\Delta G^{(2)}&=\left\langle(S_B-\left\langle S_B\right\rangle)(I_R(x_R)-\left\langle I_R(x_R)\right\rangle)\right\rangle\\\nonumber
&=\left\langle S_BI_R(x_R)\right\rangle-\left\langle S_B\right\rangle\left\langle I_R(x_R)\right\rangle.
\end{align}
Note that above method of subtracting the average item will be inaccurate in the presence of source temperature drift. Use $\frac{\left\langle S_B\right\rangle}{\left\langle S_R\right\rangle}S_R$ to replace $\left\langle S_B\right\rangle$, where $S_R=\int_{A_l}I_R(x_R)dx_{R}$, then we get
\begin{align}
G_{DGI}^{(2)}&=\left\langle(S_B-\frac{\left\langle S_B\right\rangle}{\left\langle S_R\right\rangle}S_R)(I_R(x_R)-\left \langle I_R(x_R)\right \rangle\right\rangle\\\nonumber
&=\left\langle S_BI_R(x_R)\right\rangle-\frac{\left\langle S_B\right\rangle}{\left\langle S_R\right\rangle}\left\langle S_RI_R(x_R)\right\rangle.
\label{eq:refname3}
\end{align}
This complicated form is also called DGI, it greatly improves the imaging quality compared with traditional GI. However, we must calculate all the bucket values or patterns to get the ensemble averages $\left\langle S_B\right\rangle$ and $\left\langle S_{R}\right\rangle$, which wastes calculation time.

In CI, one can conditionally average the reference signals from the positive and negative subsets, which are tactfully divided following the bucket mean or the sign of $S_B-\left\langle S_B\right\rangle$, to produce positive and negative ghost images:
\begin{equation}
\begin{cases}
G_+=\left\langle I_{R_+}\right\rangle,\ \textrm{for}\ \left\{S_{B_+}\mid S_{B}\geq\left\langle S_B\right\rangle\right\};\\
G_-=\left\langle I_{R_-}\right\rangle,\ \textrm{for}\ \left\{S_{B_-}\mid S_{B}<\left\langle S_B\right\rangle\right\}.
\end{cases}
\label{eq:refname4}
\end{equation}
Without the need to multiply the patterns by the bucket values (weights), CI magically reduces the computational complexity. Since the average value $\left\langle S_B\right\rangle$ is needed for pattern division, it requires complete measurements before calculation and still cannot realize real-time acquisition and reconstruction.

Now, to address above problems, we will present the main idea of our SGI method. First, the reference patterns and bucket values are numbered following the measurement sequence. Also starting with Eq.~(\ref{eq:refname1}), we utilize a deviation strategy here. But different from Eq.~(\ref{eq:refname2}), we use $S_{B_{i}}$ (or $I_{R_i}$) to replace $\left\langle S_B\right\rangle$ (or $\left\langle I_R(x_R)\right \rangle$). In this way, we build three modes, mode-1 for performing both successive-deviation operation, mode-2 for only deducting the adjacent bucket value, and mode-3 for only subtracting the neighboring reference pattern. For $i=1,2,\cdots,N-1$, we will have
\begin{equation}
\textrm{mode-1: }G^{(2)}_{both}=\left\langle(S_{B_{i+1}}-S_{B_i})(I_{R_{i+1}}-I_{R_i}) \right\rangle;
\label{eq:refname5}
\end{equation}
\begin{equation}
\textrm{mode-2: }
\begin{cases}
G^{(2)}_{B_+}=\left\langle(S_{B_{i+1}}-S_{B_i})I_{R_{i+1}}\right\rangle;\\
G^{(2)}_{B_-}=\left\langle(S_{B_{i+1}}-S_{B_i})I_{R_i}\right\rangle;
\end{cases}
\label{eq:refname6}
\end{equation}
\begin{equation}
\textrm{mode-3: }
\begin{cases}
G^{(2)}_{R_+}=\left\langle S_{B_{i+1}}(I_{R_{i+1}}-I_{R_i})\right\rangle;\\
G^{(2)}_{R_-}=\left\langle S_{B_i}(I_{R_{i+1}}-I_{R_i})\right\rangle.
\end{cases}
\label{eq:refname7}
\end{equation}
That is, for $N$ measurements, we can get $N-1$ pairs. Of course, we can further make the last one minus the first one to form $N$ pairs in the case of using a stable light source. As a contrast, in traditional complementary modulation schemes \cite{BSun2013,WenKai2015,Yu2015} (one pattern followed with its inverse one), $N$ measurements only generate $N/2$ effective pairs, and it is impossible to realize complementary patterns in a double-arm GI setup with rotating ground glass. Hence, our method reduces the number of measurements by half compared with complementary measurement methods. Note that here we shift one subscript to perform subtraction, we can also shift multiple subscripts for subtraction, and the corresponding results are similar. Furthermore, without the need of deducting the ensemble averages, SGI can realize real-time imaging and save computational time.

\begin{figure}[htbp]
\centering
\includegraphics[width=\linewidth]{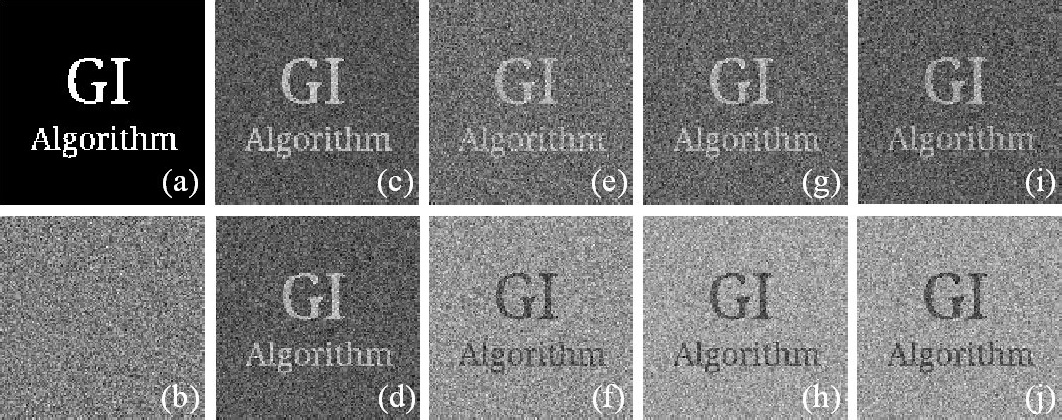}
\caption{Simulation results of $G^{(2)}$, DGI, CI, SGI mode-1, mode-2 and mode-3. (a) is the original picture of $128\times128$ pixels; (b)--(c) are the results of $G^{(2)}$ and DGI with 16,384 measurements; (d) is SGI mode-1 reconstruction. (e)--(f) are the positive and negative ghost images of CI; (g)--(h) are the retrieved results of SGI mode-2, i.e., $G^{(2)}_{B_+}$ and $G^{(2)}_{B_-}$; and (i)--(j) are the restored results of SGI mode-3, corresponding to $G^{(2)}_{R_+}$ and $G^{(2)}_{R_-}$.}
\label{fig:SimulationMethod}
\end{figure}

To demonstrate the feasibility of SGI, we create a binary image of $128\times 128$ pixels written ``GI Algorithm'' as the original image (see Fig.~\ref{fig:SimulationMethod}(a)), consisting of 0 and 1, and use random patterns of the same pixel-size for the numerical simulation of $G^{(2)}$, DGI and SGI. Figs.~\ref{fig:SimulationMethod}(b) and \ref{fig:SimulationMethod}(c) separately shows the results of $G^{(2)}$ and DGI recovered from $N=16384$ measurements. The result of SGI mode-1 is presented in Fig.~\ref{fig:SimulationMethod}(d). Obviously, SGI mode-1 can obtain a good image quality comparable to that of DGI, only by subtracting the neighboring patterns together with bucket values (rather than the ensemble averages) in traditional second-order correlation function. It is interesting to find that SGI mode-2 and mode-3 can acquire the similar phenomena of positive-negative ghost images like CI (see Figs.~\ref{fig:SimulationMethod}(e)--(f)), the corresponding results can be seen in Figs.~\ref{fig:SimulationMethod}(g)--(h) and \ref{fig:SimulationMethod}(i)--(j). Since there is no need to compute the ensemble average of bucket values for pattern division, SGI mode-2 and mode-3 can realize CI in real-time, greatly promoting the practical use of CI. The results are the same with grayscale images.

Here we introduce the contrast-to-noise ratio (CNR) as a unitless performace quantitative measure, which is defined as $\textrm{CNR}(G)\equiv\frac{\left\langle G(x_{in})\right\rangle-\left\langle G(x_{out})\right \rangle}{\sqrt{\frac{1}{2}[\Delta^{2}G(x_{in})+\Delta^{2}G(x_{out})]}}$, where $\Delta^{2}G(x)\equiv\left\langle G(x)^{2}\right\rangle-\left\langle G(x)\right \rangle^{2}$ denotes the variance, $x_{in}$ and $x_{out}$ stand for the pixel positions inside and outside the transmitting regions of the object \cite {Kam2010}. Naturally, the larger the CNR value, the better the image quality of reconstruction.

\begin{figure}[htbp]
\centering
\includegraphics[width=\linewidth]{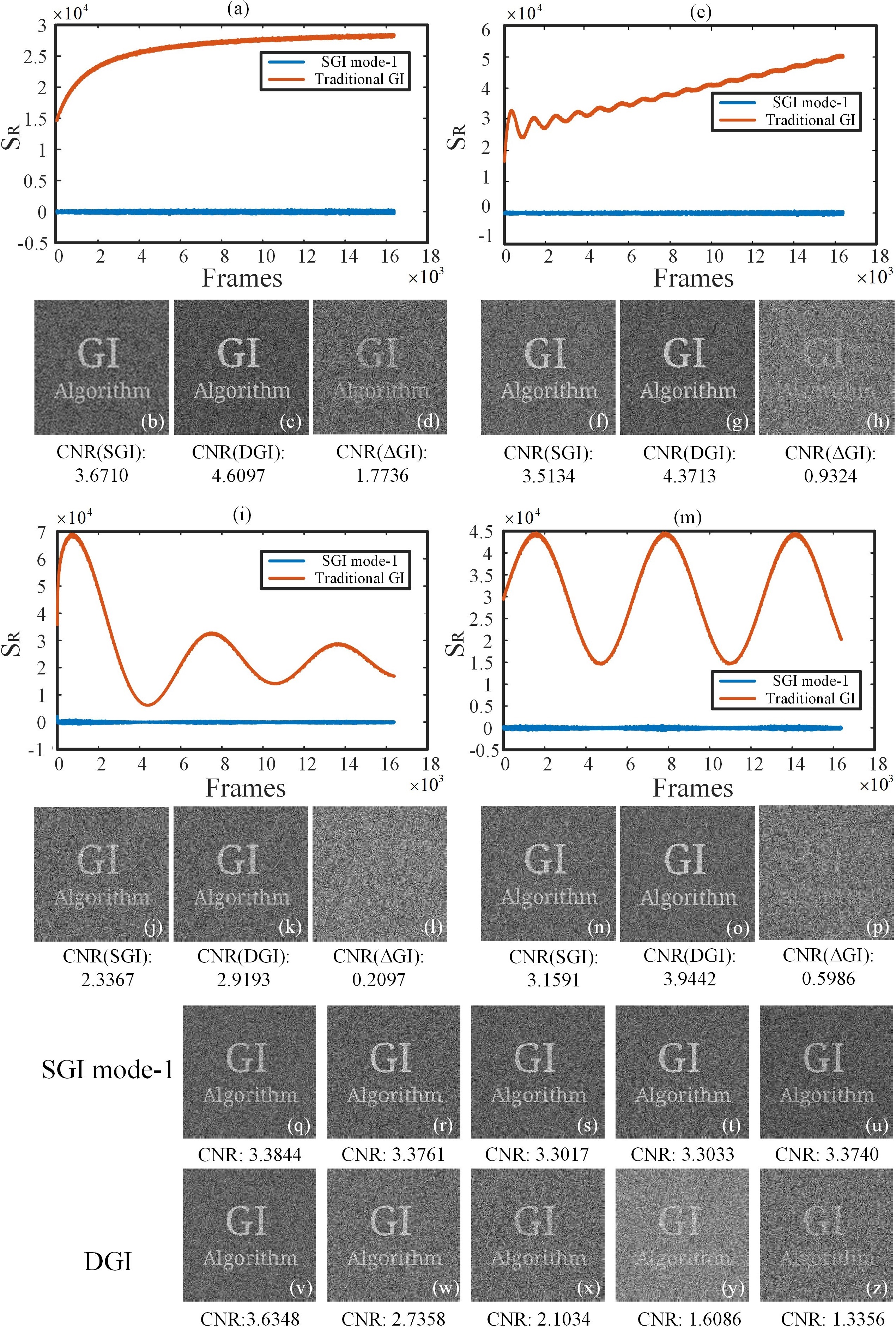}
\caption{Reconstructed results under different kinds of temperature drift of the light source and different mean values of additive white Gaussian noise. (a)--(d), (e)--(h), (i)--(l), (m)--(p) are four different kinds of temperature drifts and their related results recovered by SGI mode-1, DGI and $\Delta$GI. In the curve graphs, traditional GI involves $\Delta$GI and DGI. (q)--(u) and (v)--(z) are the restored images of DGI and SGI mode-1 under the same relatively small variance but different mean values of white Gaussian noise, with a range from 0.012 to 0.06. Note that (q)--(z) are obtained in the presence of the first kind of temperature drift, as shown in (a).}
\label{fig:DriftNoise}
\end{figure}

In order to test the robustness of SGI mode-1 against the temperature drift of light source, we simulate four kinds of $S_{R}$ fluctuations, as shown in Figs.~\ref{fig:DriftNoise}(a), \ref{fig:DriftNoise}(e), \ref{fig:DriftNoise}(i), \ref{fig:DriftNoise}(m). The ordinate $S_{R}$ denotes the total light intensity of each pattern. Obviously, SGI mode-1 can remove the effect of the source instability on $S_{R}$, while $S_{R}$ of both $\Delta$GI and DGI suffers from the temperature drift influence, showing greater fluctuations. The related reconstructions of SGI mode-1, DGI and $\Delta$GI are given in Figs.~\ref{fig:DriftNoise}(b)--(d), \ref{fig:DriftNoise}(f)--(h), \ref{fig:DriftNoise}(j)--(l) and \ref{fig:DriftNoise}(n)--(p). From these figures, we can see that SGI mode-1 and DGI have similar image qualities, better than that of $\Delta$GI in the presence of different temperature drifts. SGI mode-1 involves the adjacent pairwise subtraction of patterns, so its $S_{R}$ fluctuates slightly around zero and shows a good stability; DGI considers the influence of $S_{R}$ in the subtracted term, hence it also performs well; while $\Delta$GI ignores the effect of $S_{R}$, thus it is sensitive to the drastic fluctuations of light source. On the basis of the first kind of temperature drift, we further add white Gaussian noise of the same variance but with different mean values changing from 0.012 to 0.06 with a 0.012 stepping increase to the collecting light field. From the recovered images and CNR values as presented in Figs.~\ref{fig:DriftNoise}(q)--(z), SGI has an excellent anti-noise ability especially in the harsh detection environments with temperature drifts.
\begin{figure}[htbp]
\centering
\includegraphics[width=\linewidth]{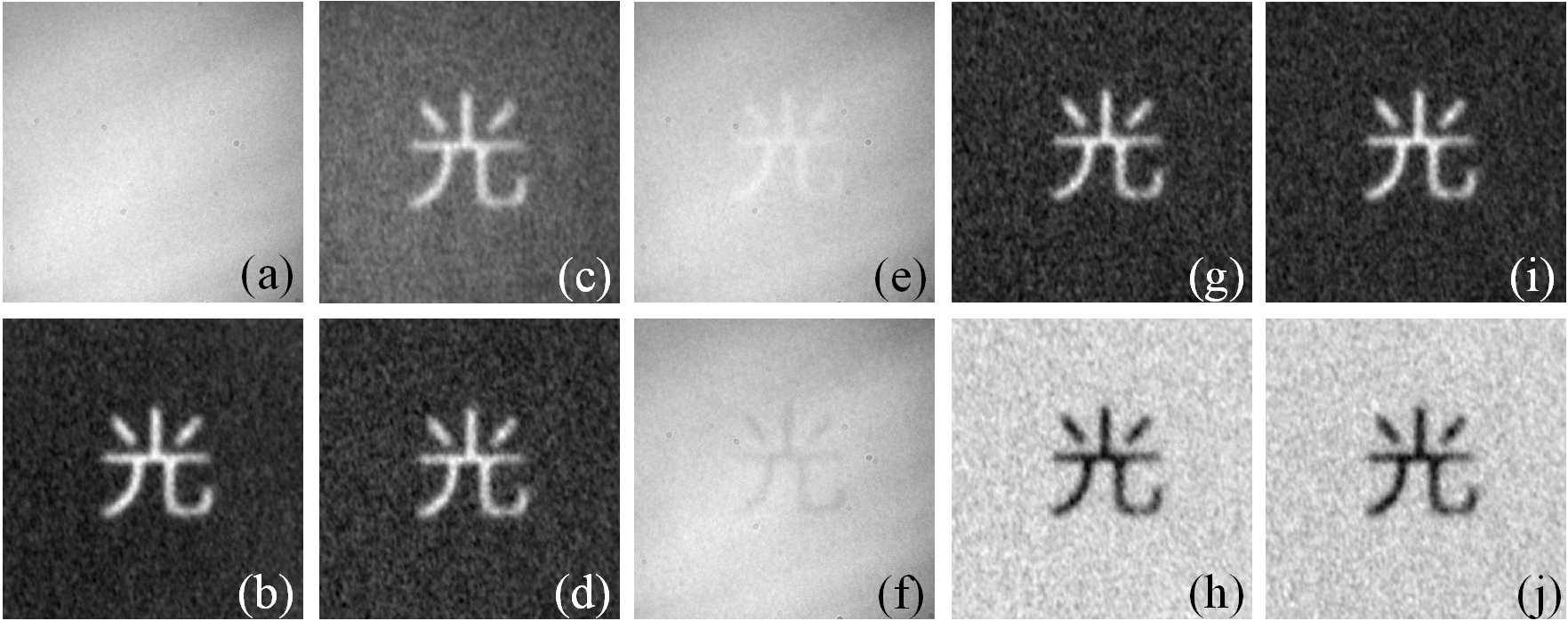}
\caption{Experimental results. (a)--(d) are the recovered results of $G^{(2)}$, DGI, $\Delta$GI and SGI mode-1; (e)--(f), (g)--(h) and (i)--(j) are the positive and negative ghost images of CI, SGI mode-2, and SGI mode-3, respectively.}
\label{fig:ExpResults}
\end{figure}

In experiments, based on the lensless GI setup shown in Fig.~\ref{fig:Setup}, we use a film printed with a Chinese character as the object to be detected, and choose a fixed imaging region of $300\times300$ pixels on the speckle patterns, thus the images recovered are also of $300\times300$ pixels. Using a high-speed camera and bucket detector with synchronization between them, we record both reference and bucket signals. In order to avoid the periodicity of the ground glass patterns, we add some disturbances/displacements in the central axis of rotation. The experimental results of $G^{(2)}$, DGI, $\Delta$GI and SGI mode-1 are shown in Figs.~\ref{fig:ExpResults}(a)--(d), and the reconstructed positive-negative images of CI, SGI mode-2 and SGI mode-3 are given in Figs.~\ref{fig:ExpResults}(e)--(f), \ref{fig:ExpResults}(g)--(h) and \ref{fig:ExpResults}(i)--(j). Here the number of patterns used is 50,000. It is experimentally demonstrated that SGI mode-1 has an outstanding performance comparable to that of DGI, which is also consistent with the simulation results. Note that the computational complexity of SGI is much smaller than that of DGI, only using simpler second-order correlation function. Additionally, it can be seen that the positive and negative images recovered by SGI mode-2 and mode-3 has superior image quality than CI, but without the need to compare each bucket values with the ensemble average, realizing real-time CI.

\begin{figure}[htbp]
\centering
\includegraphics[width=\linewidth]{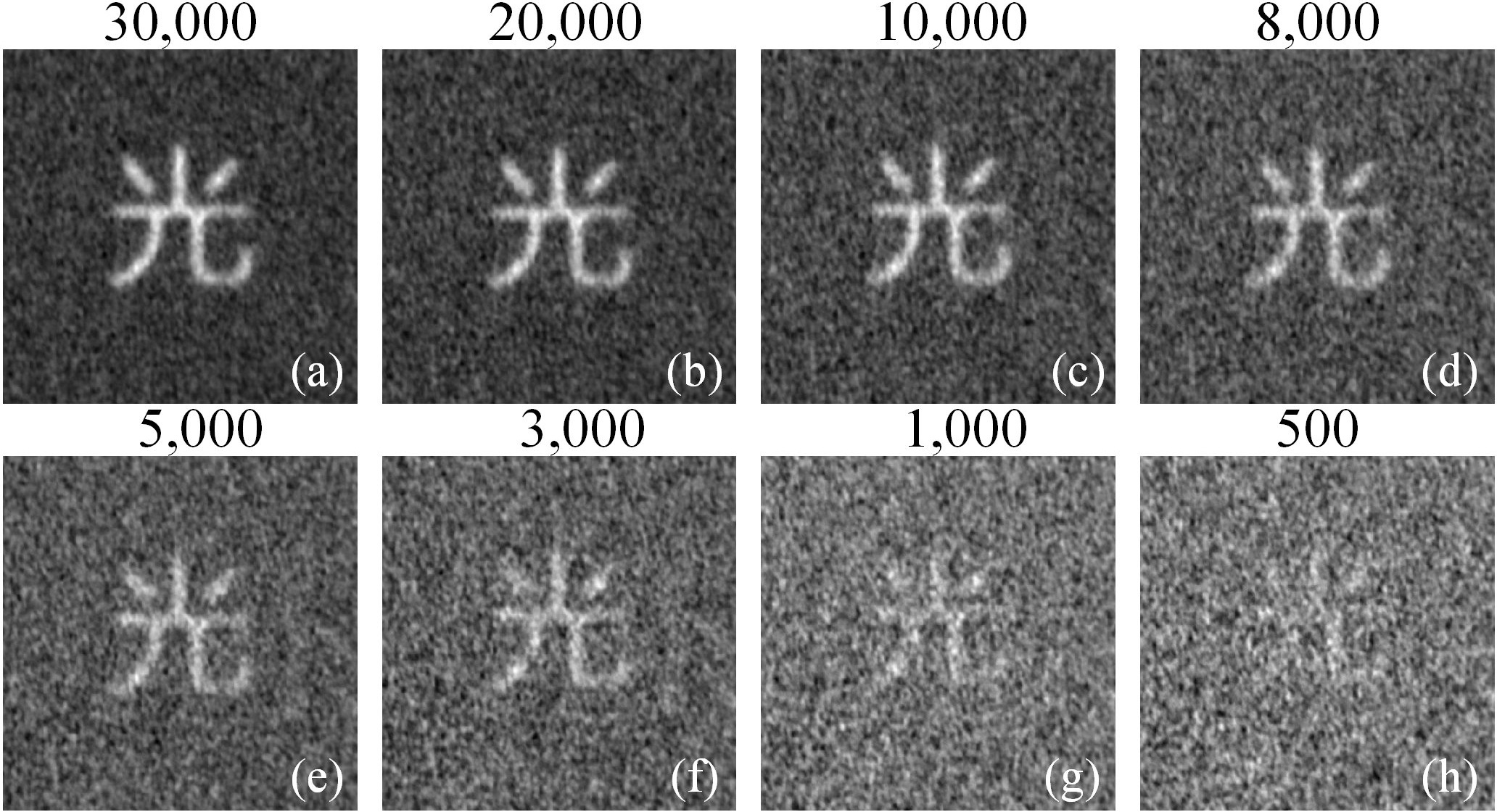}
\caption{Reconstructions of SGI from experimental data with different measurement numbers, changing from 30,000 to 500.}
\label{fig:SampleNum}
\end{figure}

Then, we gradually reduce the number of patterns from 30,000 to 500, the corresponding results of SGI mode-1 of $300\times300$ pixels are given in Fig.~\ref{fig:SampleNum}, in which the image quality is proportional to the number of measurements. When the number of patterns used equals to 5,000 (i.e, 5.6\% sampling ratio), we can get a relatively clear image, the clarity is sufficient for many practical applications. When the number of measurements is decreasing to 500 (i.e, 0.56\% sampling ratio), we can still faintly recognize the contour of the object.

Now we will offer an explanation of forming positive and negative images. It seems that CI does not need to multiply each pattern by its weight, just requires conditionally averaging partial patterns. But actually, its weights are reduced to 1 and 0 for $I_{R_+}$ (or 0 and 1 for $I_{R_-}$). This process is like converting the bucket values into a positive-negative distribution by the operations of subtracting the mean and binarization. As we know, the spatial resolution of recovered images is the same as that of patterns used, thus the focus and the main item to be averaged in second-order correlation is indeed patterns themselves. As for SGI mode-2 and mode-3, also called bucket successive-deviation and pattern successive-deviation, they make either bucket values or reference patterns minus their previous irrelevant ones. By this means, the bucket values or patterns are also turned into a positive-negative distribution, just similar to the process of CI, but without binarization. The other difference is that SGI mode-2 and mode-3 also shift the direct current background to 0 through the successive-deviation, while CI fails to remove the background noise, so the image quality of SGI mode-2 and mode-3 is better than that of CI. Note that $G^{(2)}_{both}=G^{(2)}_{B}=G^{(2)}_{R}$, where $G^{(2)}_{B}=G^{(2)}_{B_+}-G^{(2)}_{B_-}$ and $G^{(2)}_{R}=G^{(2)}_{R_+}-G^{(2)}_{R_-}$, so SGI mode-1 can be treated as a function of the two forms in SGI mode-2 and mode-3, respectively.

The role of deducting the background item in terms of $\left\langle S_B\right\rangle\left\langle I_R(x_R)\right\rangle$ or $\frac{\left\langle S_B\right\rangle}{\left\langle S_R\right\rangle}\left\langle S_RI_R(x_R)\right\rangle$ (in $\Delta G^{(2)}$ or DGI) is to shift the mean of the recovered images and to make it close to 0, the latter is better, while SGI mode-1 has the same effect with DGI. But the performance of DGI will deteriorate with the increase of mean of additive noise in the presence of temperature drift. Fortunately, SGI mode-1 offers a successive-deviation way to solve this problem, especially effective in cases of double-arm GI with a rotating ground glass for the reason that the difference between two continuous patterns are very slight.

In conclusion, we present a SGI method of three modes based on a simple domino successive-deviation strategy. Among them, SGI mode-1 can retrieve a clear image in real-time from only a few successive-differential random measurements with an image quality comparable to that of DGI, and even better in scenarios where temperature drifts of illuminating source and environmental noise coexist. It is also interesting to find that SGI mode-2 and mode-3 can realize positive-negative images, without the need to compare each bucket value with the ensemble average, thus realizing real-time high-quality CI. Besides, we also provide a brief explanation about the nature of SGI and the positive-negative ghost image phenomenon. This technology can be applied to a more severe measurement environment and promote the practical developments of GI.

\section*{\textsf{Funding}}
This work is supported by the National Natural Science Foundation of China (61801022), the Beijing Natural Science Foundation (4184098).

\bigskip





\end{document}